# In-situ fabrication of $Mo_6S_6$ nanowire terminated edges in monolayer molybdenum disulfide

Wei Huang, Xiaowei Wang, Xujing Ji, Ze Zhang, and Chuanhong Jin

State Key Laboratory of Silicon Materials, School of Materials Science and Engineering, Zhejiang University, Hangzhou 310027, China
Corresponding author: chhjin@zju.edu.cn

## ABSTRACT

Edge structures are highly relevant to the electronic, magnetic and catalytic properties of two-dimensional (2D) transition metal dichalcogenides (TMDs) and their one dimensional (1D) counterpart, i.e., nanoribbons, which should be precisely tailored for the desirable applications. In this work, we report the formation of novel $Mo_6S_6$ nanowire (NW) terminated edges in a monolayer molybdenum disulfide ($MoS_2$) via an e–beam irradiation process combined with high temperature heating in a scanning transmission electron microscope (STEM). Atomic structures of NW terminated edges and the dynamic formation process were observed experimentally. Further analysis shows that NW terminated edge could form on both Mo-zigzag (ZZ) edge and S-ZZ edge which can exhibit even higher stability superior to the pristine zigzag (ZZ) and armchair (AC) edge. In addition, the analogous edge structures can be also formed in $MoS_2$ nanoribbon and other TMDs material such as $Mo_xW_{1-x}Se_2$. We believe that the presence of these novel edge structures in 2D and 1D TMD materials may provide novel properties and new opportunities for their versatile applications including catalytic, spintronic and electronic devices.

## 1. Introduction

Atomic thin transition metal dichalcogenides (TMDs) are a class of emerging two dimensional (2D) materials with potential application in electronic [1], optoelectronic [2, 3], catalytic [4, 5] and energy storage devices [6, 7], owning to their unique structures and properties. Molybdenum disulfide ($MoS_2$) is a representative member of TMDs family which has a direct band gap ~1.8 eV when thinning down to monolayer in its H phase (trigonal prismatic coordination) [8]. So far, many efforts have been devoted to the controllable growth of large area and even wafer-scale high quality monolayer $MoS_2$ materials [9-11], 2D heterostructures [12, 13], phase and defects engineering [14-16], doping and functionalization [17-20], surface engineering [21, 22], driven by the requirement of practical devices and catalytic industry.

Like most 2D materials, monolayer $MoS_2$ has two types of edges, namely armchair (AC)-edge and zigzag (ZZ)-edge. Of them the ZZ-edge is more thermodynamically-favored which could be classified into Mo-ZZ edge or S-ZZ edge depending on the specific edge termination [23, 24]. It is well established that the catalytic activity of $MoS_2$ can mainly attribute to the presence of the edge sites while the basal plane is mostly inert during the catalytic process, in further both the number of edge sites and the edge structures play important roles in the catalytic reaction. e.g., the hydrodesulfurization (HDS) reaction, for which $MoS_2$ has been widely used over decades [25-27]. The catalytic reaction mainly take place at the ZZ edges, however, only edges with unsaturated Mo atoms or sulfur vacancies exhibit sufficient catalytic activity, i.e., the reaction activity of fourfold coordinated Mo atoms are much better than saturated sixfold Mo atoms [24, 28, 29]. In addition, varying degrees of magnetic behaviors have been theoretically predicted and experimentally verified in $MoS_2$ with different terminated edge structures, especially for its 1D system [30-32]. Typically, ferromagnetism can be observed in $MoS_2$ crystals with ZZ-terminated edge [30]. So many novel properties are associated with the terminated edges and corresponding atomic structures, however, Only limited types of edge structures have been observed experimentally in $MoS_2$ related 2D TMDs materials [23, 24, 33-36], which are far from enough for further structure and property tailoring, In this case, controllable fabrication of novel edge structures is highly desirable.

On the other hand, the rapid development of *in-situ* (S)TEM technology provide us the ability of atomic scale structural analysis along with the power of performing real-time structure modification [37-40]. It has been demonstrated to be an effective way to achieve one-dimension $M_6X_6$ (M=Mo, W; X=S, Se) metallic nanowire from monolayer TMDs sheets [41-45]. Inspired by these excellent prior works and to address the issue above, we performed an *in-situ* e-beam sculpturing under high temperature to fabricate the novel $M_6X_6$ nanowire (NW) terminated edge structures in TMDs system (including 2D crystal and one dimensional (1D) nanoribbon). These results will pave the way towards controlled fabrication of novel edge structures for TMDs materials with desirable properties.

## 2. Methods

The $MoS_2$ sample was prepared via a chemical vapor deposition (CVD) process. Molybdenum trioxide ($MoO_3$) (Aladdin, 99.9%, 1mg) powder was spread on the bottom of the quartz boat and covered with the cleaned $SiO_2$/Si substrate. The furnace was firstly heated to 600 °C in 20 min and then heated to 830 °C in 10 min, then kept for the next 10 min. The excessive sulfur source (Aladdin, 99.99%, 3g), with its temperature reaching 180 °C, was heated 2 min before the furnace temperature reached 830 °C, and then held for the next 10 min. Argon (99.999%) was used as the carrier gas, with an optimized flow rate of 20 standard-state cubic centimeter per minute (sccm) during the synthesis process. The total growth time lasted for about 10 min and then the furnace cooled down naturally.

As prepared $MoS_2$ sample was first characterized by scanning electron microscope (SEM, SU70, Hitachi, and optical microscope (Axio Imager A2m, Zeiss) (**Fig.S1**(a). Then, it was transferred onto a Micro-Electro-Mechanical System (MEMS) heating chip (Wildfire Nano-Chips XT, DENSsolutions) via a standard Polymethyl methacrylate (PMMA) assisted wet chemistry process [46]. After drying, the chip was installed into the TEM heating holder (SH70, DENSsolutions) and then loaded into electron microscope for initial characterization (**Fig. S1**(b)) and further *in-situ* fabrication. The annular dark-field (ADF-STEM) and bright field (BF-STEM) image were conducted by an aberration corrected STEM microscope (FEI Titan G2 ChemiSTEM FEI) which was operated at an acceleration voltage of 200 kV. The collection angles for the ADF imaging and BF imaging were set to 53-200 mrad and 0-10 mrad respectively. The convergence semi angles were set to be 21.4 mrad. The probe current was estimated as ~60-80 pA. The dwell time per pixel was set to 2 μs for etching (repeated scan) and 10 us for image recording. STEM images simulation were performed by using the open source software QSTEM [47].

## 3. Results and discussion

To achieve the desirable edge structures, we first created fresh edges by forming holes in the pristine monolayer $MoS_2$ through an e-beam sculpting process at high temperature condition. As prepared $MoS_2$ sample was heated to 750 °C, at which temperature an atomic flat zigzag edge can be produced by e-beam etching and thermal effect [36]. Under this temperature, an ultra-clean sample area sized up to micron scale could be achieved. To enhance the etching efficiency, electron dose rate was first increased to ~300 A/cm$^2$ to form the initial hole and more moderate condition (repeated scan mode) was used in the subsequent etching process (electron dose rate was decreased to ~12 A/cm$^2$). In this way we can fabricate holes with uniform but tunable size, which usually exhibit hexagonal or truncated triangular morphology (as shown in **Fig. 1**(a-b)). ADF-STEM image in **Fig. 1**(a) shows the typical edge structures of the holes created by e-beam irradiation and high temperature heating at 750 °C. It is obvious that the created edges are strictly along the ZZ direction with atomic flat surface, which is similar to the results achieved from low-voltage e-beam etching in high temperature [36], and no AC edge was observed. It was observed that the S ZZ edges (marked with orange line) were always dominating, which are due to the surface-dependent etching rate, as the average etching rate of Mo ZZ edge (marked with blue line) is much faster than S ZZ edge (Movie S1). It is noteworthy that during the e-beam etching process, sulfur atoms are first removed by the knock on effect due to the low threshold for removing from the edge [42], leaving the Mo atoms which tend to aggregate and form adsorbed clusters with different size (marked by the dotted circle in **Fig. 1**(b)). For few layer Mo adsorbed clusters, hexagonal coordination are mostly favored, typically, for monolayer adsorbed atoms, Mo atoms tends to adsorbed on the Mo top sites. For the thick clusters, it tends to form the original body-centered cubic (bcc) structures. The detail structures of Mo clusters with different size was shown in **Fig. S2**.

By performing continued irradiation at the same temperature condition of 750 °C, abnormal contrast was frequently observed on the edge of these etched holes as the ADF-STEM image shown in **Fig. 1**(b) (see the white arrow). Then we amplified a random selected holes which marked by the dotted rectangular box, as **Fig. 1**(c) shown, atomic sharp linear structures formed on both Mo ZZ edge and S ZZ edge, which nearly cover the whole edges of the hole. To confirm the detailed structure of these edges, atomic scale STEM analysis along both S-ZZ edge (**Fig. 2**(a-b)) and Mo-ZZ edge (**Fig. 2**(d-e)) were preformed respectively. For S-ZZ edges, As the ADF-STEM in **Fig. 2**(a) and BF-STEM image in **Fig. 2**(b) shown, newly formed edge structure (marked by the rectangular box in **Fig. 2**(a)) on the S-ZZ edge can be identified as a $Mo_6S_6$ nanowire structure (Corresponding atomics structural model is presented in **Fig. 2**(c).). Generally the rotation angle (α, β) can be used to describe the rotation state of $M_6X_6$ nanowires [41], while α represents out-of-plane rotation angle and β represents rotation angle along its own c-axis. In the similar way as formed nanowire on S ZZ edge can be ascribed as (~0º, ~30º). We can further prove that the $Mo_6S_6$ nanowire are not adsorbing on the top of $MoS_2$ sheet but somehow bonding with S-ZZ edge in $MoS_2$ sheet with coherent interface with the help of STEM simulation (**Fig. S3**), thus we can ascribe it as a new kind of terminated edge. It is noteworthy that we also observe a (~0º, ~0º) NW edge on the S-ZZ edge (**Fig. S**4), although it is not as common as (~0 º, ~30º). In terms of Mo ZZ edge, As the ADF-STEM in **Fig. 2**(d) and BF-STEM image in **Fig. 2**(e) shown, just as the S edge, a $Mo_6S_6$ terminated edge formed on its outer surface with (~0º, ~0º) rotation angle (Corresponding atomics structural model is presented in **Fig. 2**(f)). Typically, for the Mo-ZZ-NW edge, all the case we observed are (~0º, ~0º), no exception was found. We also noticed the distance of the Mo-Mo columns in $MoS_2$ sheet near the NW edge on both S-ZZ edge (marked by the dotted rectangular box) and Mo-ZZ edge are compressed to 0.21 nm, compared to the value of 0.27 nm in the bulk region. The similar compression can be observed in S-DT edge [34] and reconstructed Mo-ZZ edge [33], In particular, the compression occurred on the Mo-ZZ edge may induced by the sulfur vacancies. Since the pristine edges we achieved are all along ZZ direction, no NW terminated edges formed on AC edges were observed.

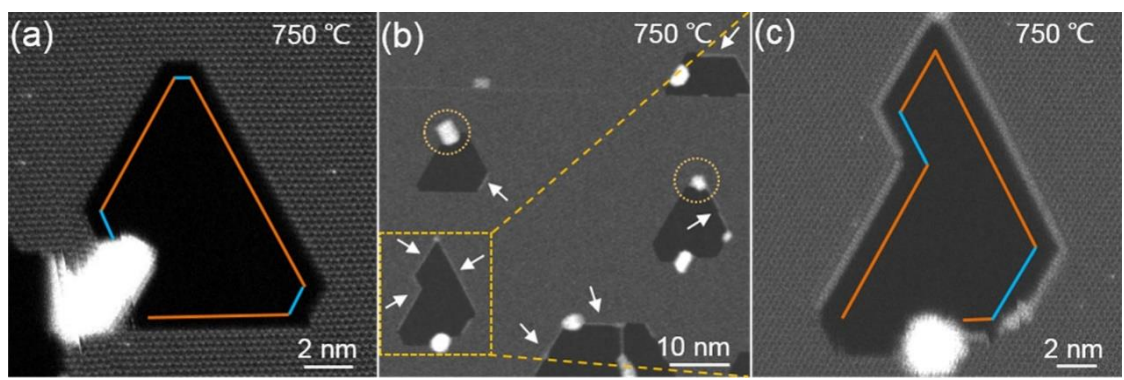

Fig. 1 Edge structures of the etching holes in a monolayer $MoS_2$ achieved at 750 ºC. (a) ADF-STEM image (Wiener filter) of the fresh zigzag edges created by e-beam etching process. The blue line represents the Mo ZZ edges and the orange line represents the S ZZ edges. (b) ADF-STEM image of the edges with abnormal contrast (marked by the white arrow) which formed by performing continued e-beam irradiation. Adsorbed Mo clusters are marked by the dotted circles. (c) Amplified ADF-STEM image of the area marked with dotted rectangular box in Fig. 1(b).

After $Mo_6S_6$ NW terminated edge formed, it can exhibit high stability under the continued e-beam irradiation, which can also preserve the inner part of $MoS_2$ sheet from further etching, while the nearby edge without the preservation of NW terminated edge, will occur the etching process. We also did a cooling test to verity its stability on relative low temperature, we found as formed NW edge can keep its structure from room temperature to at least 750 ºC, one possible explanation to the ultra-stability of NW edge is that the dangling bonds in the pristine edge can be largely reduced by forming the NW edge, which may enhance the edge stability. Besides the observed vacuum stability, a good ambient stability of $Mo_6S_6$ NW edge could also be expected since the independent $Mo_6X_6$ (X=S or Se) nanowire can be very stable in the atmosphere [48, 49]. Basically, the NW terminated edges can cover the whole edges of the hole, thus, the length of the NW terminated edges are tunable by simply creating holes with variable sizes. So far, the longest NW terminated edge we observed is 19.3 nm, the attainable size could be even longer.

A typical formation process of $Mo_6S_6$ NW terminated edges along ZZ direction was captured, which can help us understand its formation mechanism. We kept the irradiation and temperature conditions consist with the etching process (temperature was maintained at 750 ºC and the dose rate was cal. ~12 A/cm$^2$). A corresponding time-sequential ADF-STEM images was shown in **Fig. 3**(a). At t= 0 s, a few layer Mo cluster which produced by the e-beam irradiation was absorbed on the monolayer $MoS_2$ near the ZZ edge, after ~0.52 s, the adsorbed Mo clusters migrated and rearranged along the ZZ edge driven by the heating and e-beam effect [50]. Immediately the adsorbed Mo atoms combined with the edge atoms of $MoS_2$ experienced a complicated structural reconstruction to form the initial NW terminated edge nucleation in another ~0.52 s, we propose the growth rate is not reaction controlled as the reconstruction process is very fast, which exceed the time resolution limit (0.52 s). Based on our experiment, in this step, an average growth rate could be estimated at the level of ~10 nm/s. After that, as formed nucleus absorbed atoms of nearby edge from etching process and grew longer to form the stable NW terminated edge structures, In this step, the growth rate was restricted by the etching rate of nearby edges. An average growth rate could be estimated at the level of 0.1 nm/s. Furthermore, we discuss the alternative route towards ZZ-NW terminated edge structures as observed in our experiments. Take Mo edge for example, scheme illustration in **Fig. 3**(b) shows two kinds of simplified routes towards NW terminated edge. Route I is much closed to the case we discussed above, which the formation of $Mo_6S_6$ NW terminated edge is assisted by the high activity few layer adsorbed Mo cluster. The chemical reaction could be written as Mo (cluster) + $MoS_2$ → $Mo_6S_6$ (nanowire). Route II is another common case. In the first step, Sulfur atoms of the edge are first removed by the e-beam and the remaining Mo atoms directly adsorbed along the edge, then the adsorbed structure undergo the same structural reconstruction just as route I. The chemical reaction could be written as (1) $MoS_2$ → Mo (adsorbed atom) + S, (2) Mo (adsorbed atom) + $MoS_2$ → $Mo_6S_6$ (nanowire). Considering that the Mo cluster in route I is also come from the e-beam irradiation, two routes are actual equivalent. However, limited to the irreconcilable contradiction between the signal to noise ratio and time resolution for the STEM scanning, the detailed transition process of $Mo_6S_6$ NW edge can't be directly resolved, which need further consideration. It is worthwhile mentioning that a Mo rich or S deficiency environments are always favorable for the formation of NW terminated edge structures as nanowires are S deficiency structure compared to the $MoS_2$ sheet, while this condition can be easily created in TEM or STEM with the help of e-beam irradiation.

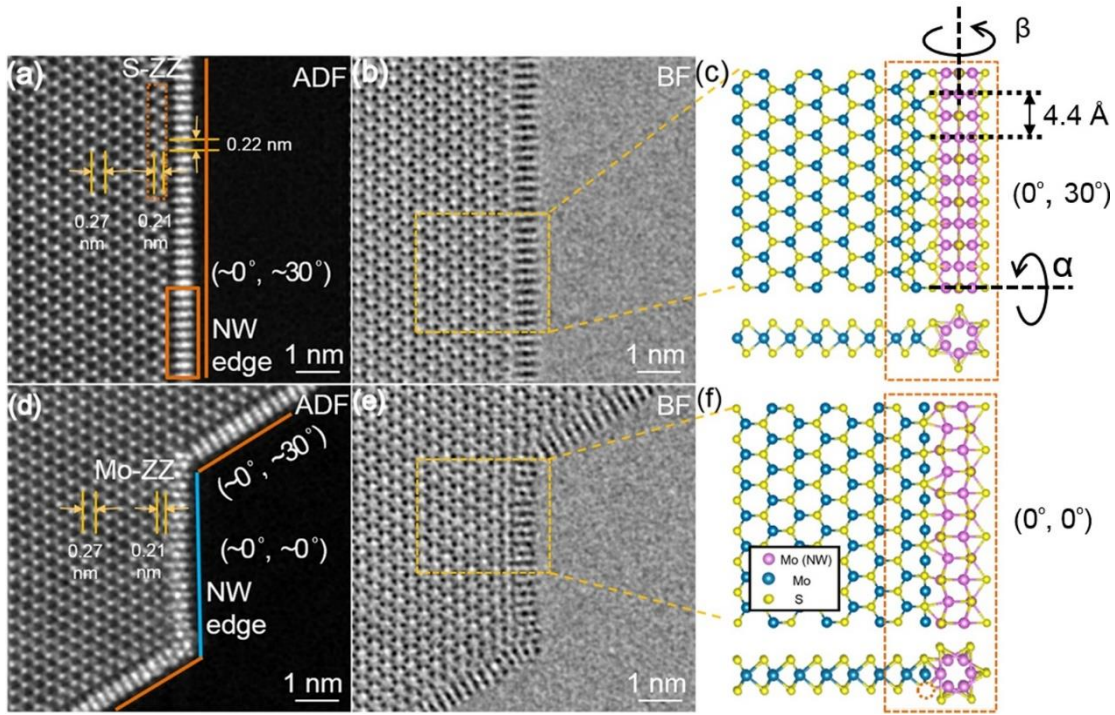

Fig. 2 Atomic structures of ZZ-$Mo_6S_6$ NW terminated edges. (a) Atomic resolution ADF-STEM image (Wiener filter) of (~0 °, ~30 °) S-ZZ-NW terminated edge (marked by the rectangular box), the compression region in $MoS_2$ sheet was marked by the dotted rectangular box, with Mo-Mo distance compressed to 0.21 nm, (b) corresponding BF-STEM image (Wiener filter) and (c) atomic structural models marked by the dotted rectangular box in Fig. 2b. (d) Atomic resolution ADF-STEM image (Wiener filter) of (~0 °, ~0 °) Mo-ZZ-NW terminated edge, (e) corresponding BF-STEM image (Wiener filter) and (f) atomic structure models marked by the dotted rectangular box in Fig. 2e, dotted circle shows the existence of sulfur vacancies on the compressed Mo ZZ edge.

Besides the e-beam effect, preparation temperature is another key factor for controllable fabrication of $Mo_6S_6$ NW edge. Here we explain why high temperature heating is essential to form NW terminated edge structures. **Fig.S5(a-b)** shows the typical edge structure achieved at 400 °C with different e-beam exposure time (dose rate was kept at ~12 A/cm$^2$). At this temperature, we didn't get the same $Mo_6S_6$ NW edge which can be easily achieved at 750 °C, but rough edge with random stacking Mo clusters and absorbed/suspending $Mo_6S_6$ nanowires. More severe beam damage and sample contamination under e-beam irradiation makes the preparation of NW edge at a relative low temperature even harder. Combined with the results achieved under other temperature conditions, we supposed the ideal temperature used to fabricate NW terminated edge should be around 750 °C.

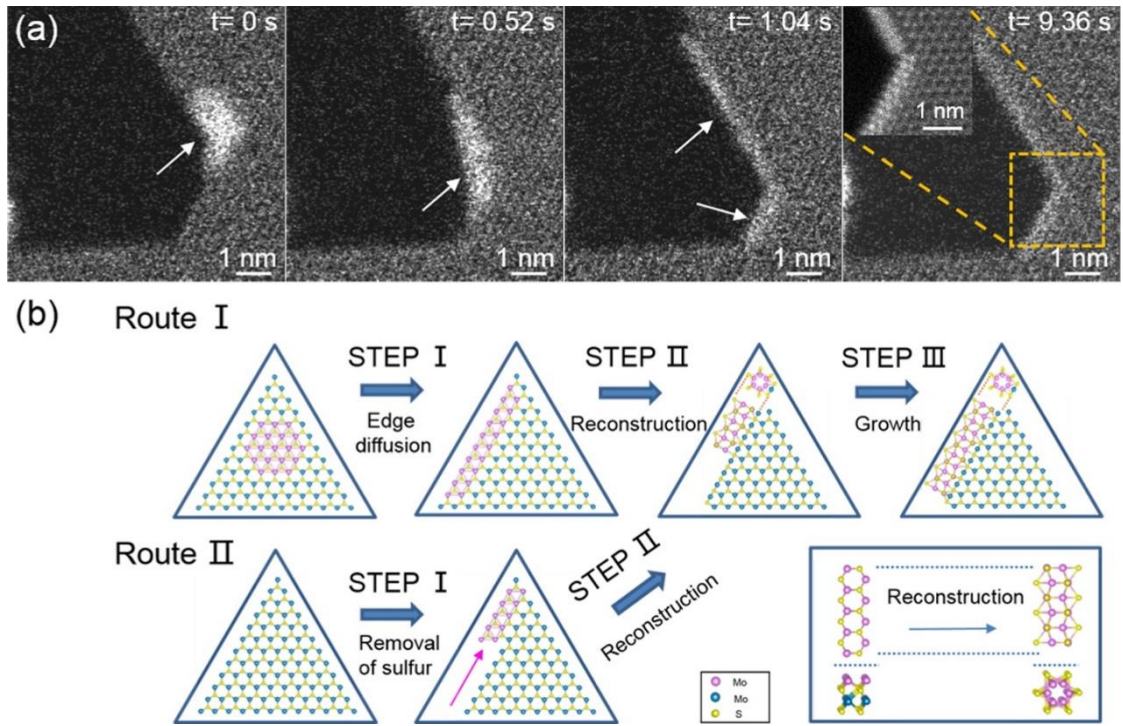

Fig.3 Dynamic formation process of ZZ-$Mo_6S_6$ NW terminated edges (a) Time-resolution ADF-STEM images of the dynamic formation process of NW terminated edge, which was assisted by few layer adsorbed Mo cluster. (b) Scheme illustration of alternative routes towards NW terminated edge and corresponding simplified model of structural transformation from Mo adsorbed structure to $Mo_6S_6$ edge structure.

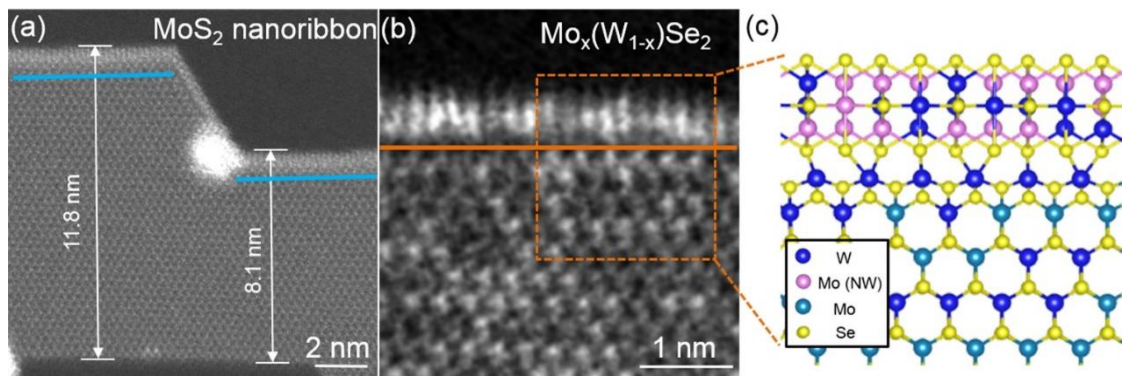

Fig. 4 ZZ-NW terminated edges in 1D $MoS_2$ nanoribbon and $Mo_xW_{1-x}Se_2$. (a) ADF-STEM image of an in-situ fabricated $MoS_2$ nanoribbon with ZZ-NW edge on one side. Blue line represents the Mo-ZZ-NW edge while the other side is the S-DT edge. (b) Observed ZZ-NW terminated edge formed on Se edge in $Mo_xW_{1-x}Se_2$ material.

On the other hand, dimension reduction from 2D to 1D structure can magnify the effect of edge states [51]. In this case, edge structure tailoring in 1D system is more essential. By using the similar *in-situ* method, we also fabricate a nanoribbon with NW terminated edge on one side (mainly along Mo ZZ edge, marked by the blue line), while the other side is S DT edge (**Fig.4** (a)). The nanoribbon with NW terminated is size adjustable as the widest part is 11.8 nm and the thinnest part is 8.1 nm. It is natural to expect that the existence of other types of nanoribbon with NW terminated edges, e.g. $MoS_2$ nanoribbon with NW terminated edges on both sides; one side with Mo-ZZ edge and the other side with NW terminated edge. Our findings suggest that the similar method to produce NW terminated edge is applicable for 1D system and provide more possibility for further tailoring. It is important to highlight that the NW terminated edge can also be observed in $Mo_xW_{1-x}Se_2$ system (**Fig. 4**(b)), which indicate that results from $MoS_2$ sheets are highly portability and can extend to other TMD materials. Finally, we suppose the fabrication of NW terminated edges may have variety of meanings: (i) the $Mo_6S_6$ nanowire has been reported exhibiting a metallic behavior, while the $MoS_2$ sheet is semiconductor, the metal-semiconductor interface may introduce novel physics and properties. (ii) High density unsaturated Mo atoms at NW edge along with the extremely stability may enhance the catalytic behavior. (iii) The metallic nanowire may reduce the contact resistance between the semiconductor and metal-wires in the real electronic devices. (iiii) A good system to study 1D electron dynamics, e.g., 1D charge density waves [52, 53].

## 4. Conclusions

In conclusion, a novel type of $Mo_6S_6$ NW edge structure was fabricated via an in-situ TEM methods. The detail atomic structures were examined and the formation process was captured. All $Mo_6S_6$ NW terminated edges we created are along ZZ direction, no AC-NW edge were observed. Atomic resolution STEM images reveal nanowires bonding with different zigzag edge usually shows selective rotation angle. Typically for the Mo ZZ edge it tends to form stable (~0º, ~0º) structure and more degree of freedom for the S ZZ edge as (~0º, ~30º) and (~0º, ~0º) were observed. To achieve the desire edge structures, we suppose that a Mo rich or S deficiency environment are essential, which can be created by e-beam. High temperature conditions are also play an important role while the best condition is around 750 ºC based on our experiments. The ultra-stability of the NW terminated edge with wide temperature range as well as high portability in the $MoS_2$ nanoribbon and other TMDs materials like $Mo_xW_{1-x}Se_2$ system lights up the way for the potential application. Our results present here will encourage more theoretical study and experimental exploration on these novel edge structures and properties.

## Acknowledgements

This work was financially supported by the National Science Foundation of China under Grants 51772265, 51472215, 5171165024 and 61721005, the National Basic Research Program of China under Grants 2014CB932500 and the 111 project under Grant B16042. The work on electron microscopy was done at the Center of Electron Microscopy of Zhejiang University. We also thank Wandong Gao and Chuqiao Shi for the help of CVD sample preparation.

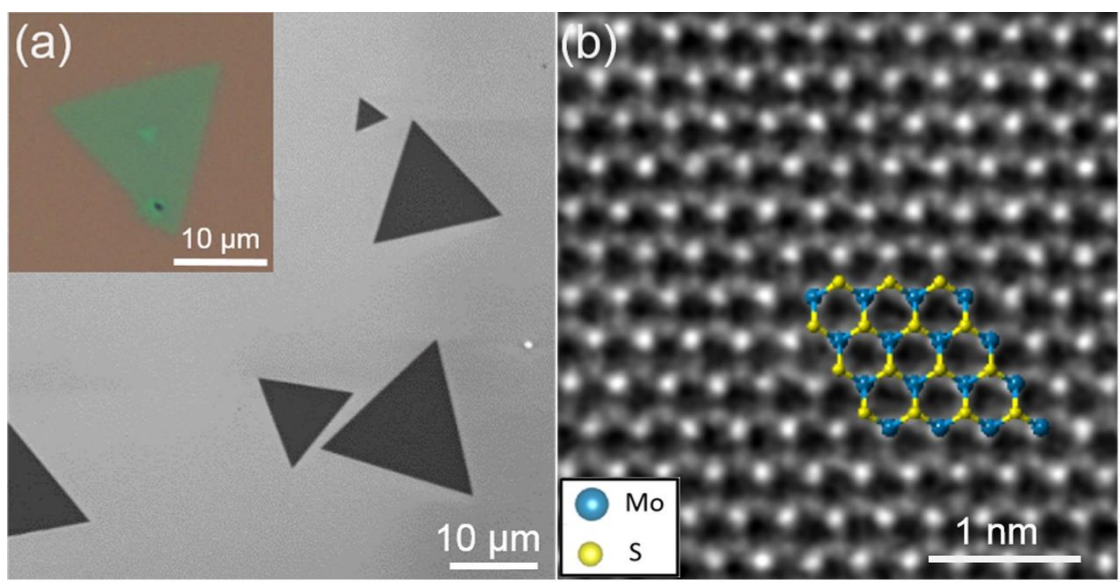

Fig. S1 characterization of pristine MoS$_2$ sample via a typical CVD process. (a) SEM image of as prepared MoS$_2$ sample and corresponding optical image (inset), which shows the representative triangular-shape morphology of our CVD samples, the size of which ranges from several micrometers to dozens of micrometers. (b) High resolution ADF-STEM image (Wiener filter) of a monolayer MoS$_2$ with atomic structure model overlaid. The brighter spots represent the Mo atoms and the darker spots represent the S atoms which is consistent with the typical hexagonal symmetry of a monolayer 2H-MoS$_2$. The ADF-STEM image also reflects the high crystallinity of the sample.

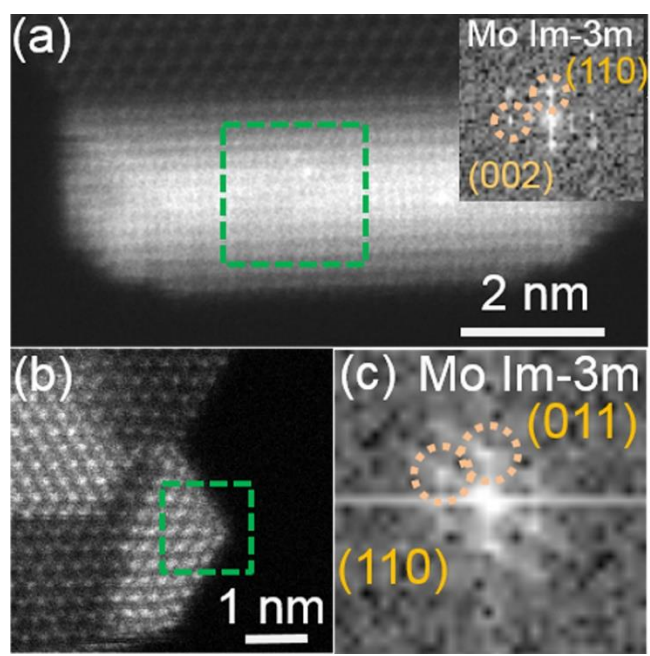

Fig. S2 Atomic resolution ADF-STEM images of the Mo cluster adsorbed on the edge with different projection direction and corresponding Fast Fourier Transformation (FFT). (a) Mo cluster projected along the [-110] direction, the inset is corresponding FFT of the area marked by the dotted rectangular box. (b) Few layer Mo cluster projected along the [1-11] direction, (c) corresponding FFT of the area marked by the dotted rectangular box in Fig.S2 (b).

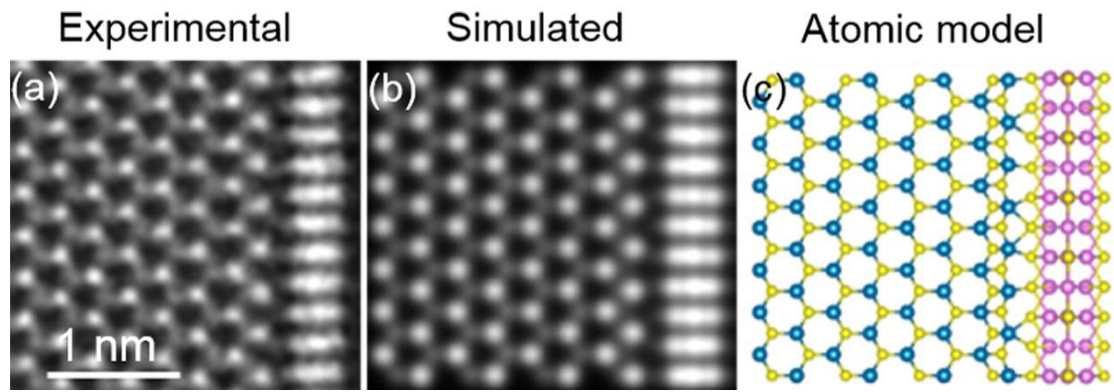

Fig. S3 STEM simulation of S-ZZ-NW edge. (a) Experimental ADF-STEM image (Wiener filter). (b) Simulated image. (c) Atomic structural model.

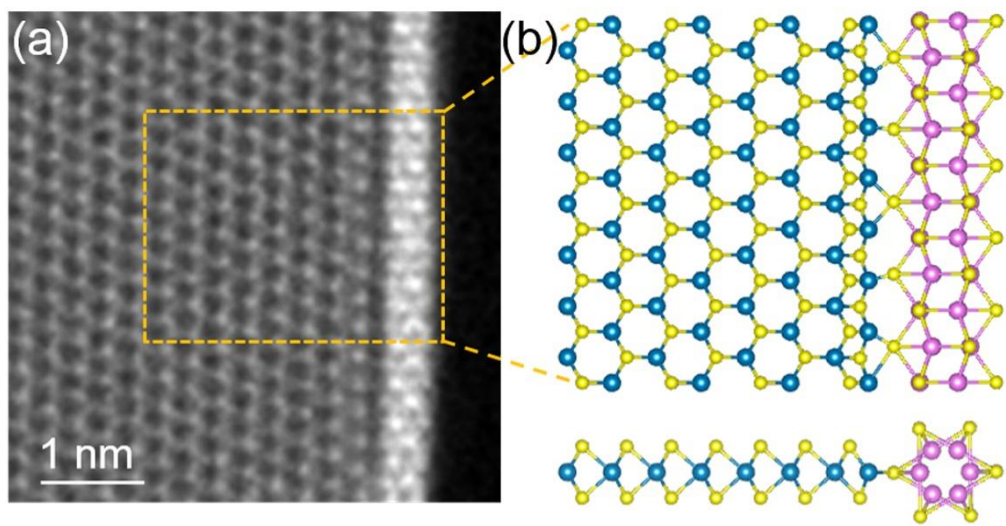

Fig. S4 Atomic structure of (~0 °, ~0 °) S-ZZ-NW edge. (a) ADF-STEM image (low-pass filtered image). (b) Corresponding atomic structural model of the area marked by the dotted rectangular box in Fig. S4 (a).

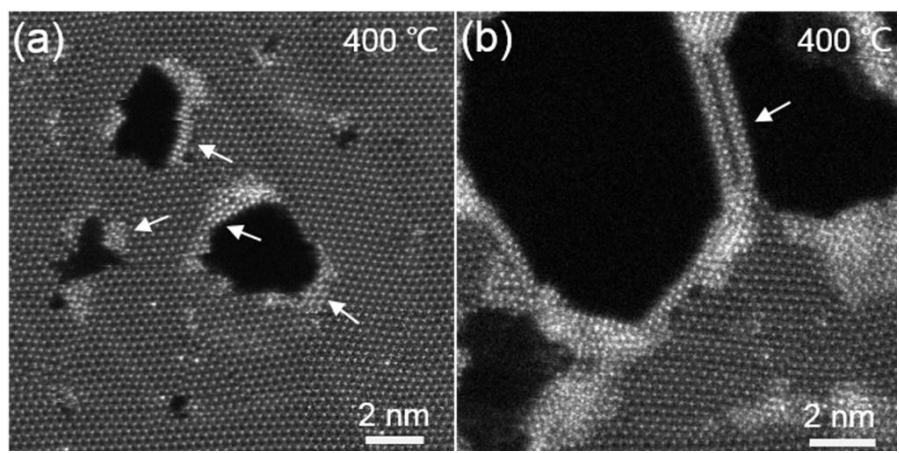

Fig. S5 Edge structures of the hole achieved at 400 °C. (a) ADF-STEM image of the edge structure obtained by exposing the sample to e-beam for a few seconds. (b) Edge structures achieved by performing continued e-beam irradiation, suspended nanowires was marked by the white arrow.